\documentstyle[12pt,epsfig]{article}
\newcommand{\be}{\begin{equation}}
\newcommand{\ee}{\end{equation}}
\newcommand{\beq}{\begin{equation}}
\newcommand{\eeq}{\end{equation}}
\newcommand{\ba}{\begin{eqnarray}}
\newcommand{\ea}{\end{eqnarray}}
\newcommand{\bea}{\begin{eqnarray}}
\newcommand{\eea}{\end{eqnarray}}

\begin{document}
\baselineskip=15.5pt \pagestyle{plain} \setcounter{page}{1}
%--------+---------+---------+---------+---------+---------+---------+
%Body

% Ofer's definitions

\def\del{{\partial}}
\def\vev#1{\left\langle #1 \right\rangle}
\def\cn{{\cal N}}
\def\co{{\cal O}}
\newfont{\Bbb}{msbm10 scaled 1200}     %instead of eusb10
\newcommand{\mathbb}[1]{\mbox{\Bbb #1}}
\def\IC{{\mathbb C}}
\def\IR{{\mathbb R}}
\def\IZ{{\mathbb Z}}
\def\RP{{\bf RP}}
\def\CP{{\bf CP}}
\def\Poincare{{Poincar\'e }}
\def\tr{{\rm tr}}
\def\tp{{\tilde \Phi}}

\def\TL{\hfil$\displaystyle{##}$}
\def\TR{$\displaystyle{{}##}$\hfil}
\def\TC{\hfil$\displaystyle{##}$\hfil}
\def\TT{\hbox{##}}
\def\HLINE{\noalign{\vskip1\jot}\hline\noalign{\vskip1\jot}} %Only in latex
\def\seqalign#1#2{\vcenter{\openup1\jot
  \halign{\strut #1\cr #2 \cr}}}
\def\lbldef#1#2{\expandafter\gdef\csname #1\endcsname {#2}}
\def\eqn#1#2{\lbldef{#1}{(\ref{#1})}%
\begin{equation} #2 \label{#1} \end{equation}}
\def\eqalign#1{\vcenter{\openup1\jot
    \halign{\strut\span\TL & \span\TR\cr #1 \cr
   }}}
\def\eno#1{(\ref{#1})}
\def\href#1#2{#2}
\def\half{{1 \over 2}}

%--------+---------+---------+---------+---------+---------+---------+
%Hirosi's macros:
\def\ads{{\it AdS}}
\def\adsp{{\it AdS}$_{p+2}$}
\def\cft{{\it CFT}}

\newcommand{\ber}{\begin{eqnarray}}
\newcommand{\eer}{\end{eqnarray}}

\newcommand{\beqar}{\begin{eqnarray}}
\newcommand{\cN}{{\cal N}}
\newcommand{\cO}{{\cal O}}
\newcommand{\cA}{{\cal A}}
\newcommand{\cT}{{\cal T}}
\newcommand{\cF}{{\cal F}}
\newcommand{\cC}{{\cal C}}
\newcommand{\cR}{{\cal R}}
\newcommand{\cW}{{\cal W}}
\newcommand{\eeqar}{\end{eqnarray}}
\newcommand{\th}{\theta}
\newcommand{\lm}{\lambda}\newcommand{\Lm}{\Lambda}
\newcommand{\eps}{\epsilon}
\newcommand{\pa}{\paragraph}
\newcommand{\pt}{\partial}
\newcommand{\de}{\delta}
\newcommand{\De}{\Delta}
\newcommand{\lb}{\label}

%--------+---------+---------+---------+---------+---------+---------+

\newcommand{\nonu}{\nonumber}
\newcommand{\oh}{\displaystyle{\frac{1}{2}}}
\newcommand{\dsl}
  {\kern.06em\hbox{\raise.15ex\hbox{$/$}\kern-.56em\hbox{$\partial$}}}
\newcommand{\id}{i\!\!\not\!\partial}
\newcommand{\as}{\not\!\! A}
\newcommand{\ps}{\not\! p}
\newcommand{\ks}{\not\! k}
\newcommand{\D}{{\cal{D}}}
\newcommand{\dv}{d^2x}
\newcommand{\Z}{{\cal Z}}
\newcommand{\N}{{\cal N}}
\newcommand{\Dsl}{\not\!\! D}
\newcommand{\Bsl}{\not\!\! B}
\newcommand{\Psl}{\not\!\! P}
\newcommand{\eeqarr}{\end{eqnarray}}
\newcommand{\ZZ}{{\rm \kern 0.275em Z \kern -0.92em Z}\;}

\begin{titlepage}

%\leftline{\tt hep-th/yymmnnn}

\vskip -.8cm

\begin{center}

\vskip 1.5 cm

{\LARGE Kerr-AdS black holes and non-relativistic conformal QM theories
in diverse dimensions} \vskip .3cm

\vskip 1.cm

{\large Martin Schvellinger}

\vskip 0.6cm
{\it IFLP-CCT-La Plata, CONICET and \\
Departamento de F\'{\i}sica, Universidad Nacional de La Plata
\\ Calle 49 y 115, C.C. 67, (1900) La Plata, \\ Buenos Aires, Argentina}

\vspace{1.7cm}

{\bf Abstract}

\end{center}

We study the discrete light cone quantization of Kerr-AdS$_{d+3}$ black holes with
a plane wave boundary. The resulting backgrounds
are conjectured to be dual of non-relativist conformal quantum mechanical
theories in $d$ spatial dimensions at finite temperature and finite chemical potential.
This includes the discrete light cone quantization of Kerr-AdS$_4$
and Kerr-AdS$_6$ black holes. These are conjectured to be dual of non-relativistic
conformal quantum mechanical theories in one and three spatial dimensions, respectively,
which are defined on plane wave backgrounds. We also consider the case of the BTZ black hole.
We calculate thermodynamic properties of these systems by using the gravity dual models.
We discuss the embedding of these backgrounds in string theory and M-theory.
We give general formulas for the discrete light cone quantization of Kerr-AdS
black holes in higher dimensions and their thermodynamic properties.

\noindent

\end{titlepage}

\newpage

\tableofcontents

\newpage
%--------+---------+---------+---------+---------+---------+---------+
%Body

\vfill

\section{Introduction}

In the last few years the AdS/CFT correspondence \cite{Maldacena:1997re,Gubser:1998bc,Witten:1998qj}
has been applied to modeling the dynamics of strongly coupled field theories which describe
diverse physical systems in a very broad context. For instance, it has been used to model
properties of the strongly coupled dynamics of quark-gluon plasmas, including transport properties.
Also it has become a worth tool to explore the strongly coupled dynamics of certain condense
matter systems.

The AdS/CFT correspondence provides a dictionary to map relativistic quantum field
theories to the corresponding string theory or gravity dual systems.
Starting from a relativistic quantum mechanical theory and considering its discrete
light cone quantization (DLCQ), a non-relativistic quantum mechanical theory is obtained.
If the generators of the conformal group are included, the resulting theory
will become a non-relativistic conformal quantum mechanical (NRCQM) relative, which
will have the symmetry generated by the Schr\"odinger group.
Very recently, NRCQM theories obtained by performing DLCQ of
certain field theories as well as their dual gravity backgrounds have been studied
\cite{Maldacena:2008wh,Adams:2008wt,Herzog:2008wg}. In these
references string theory solutions with non-relativistic conformal symmetry have also been considered.

The interest in the investigation of gravity solutions with non-relativistic conformal
symmetry has recently grown, motivated by the expectation that they would be the
dual description of conformal quantum mechanical systems, focussed upon applications to certain
condense matter problems \cite{Herzog:2007ij,Hartnoll:2007ih,Hartnoll:2007ai,Hartnoll:2007ip,
Gubser:2008px,Hartnoll:2008vx,Son:2008ye,Balasubramanian:2008dm,Wen:2008pb,Sakaguchi:2008rx,
Gubser:2008wv,Roberts:2008ns,Jejjala:2008jy,Goldberger:2008vg,Barbon:2008bg,Sakaguchi:2008ku,
Wen:2008hi,Minic:2008xa,Chen:2008ad,Schafer:2008nv,O'Bannon:2008aj,Galajinsky:2008ig,Kachru:2008yh,Davis:2008nv,
Kovtun:2008qy,Leiva:2003kd,Duval:2008jg,Lee:2008xf,Yamada:2008if,Lin:2008pi,Hartnoll:2008rs,Hartnoll:2008kx}.
If the NRCQM
theory admits a gravity dual description, it would be expected that the DLCQ of a certain string theory or M-theory
background will be its gravity dual. The examples worked out in \cite{Maldacena:2008wh} include the
cases of metrics which asymptotically approach AdS spaces in five and seven dimensions. They
are conjectured to be dual of NRCQM theories in two and four spatial dimensions, respectively.
So, it is very interesting
to extend the analysis to backgrounds whose asymptotic limits are AdS spaces in diverse dimensions,
including those which should render dual NRCQM theories in one and three spatial dimensions.
Hence, in this paper we consider such extensions, based upon the proposal of reference  \cite{Maldacena:2008wh}.
 Also we calculate thermodynamic
properties of the gravity dual systems in $d+3$ dimensions which should be the corresponding thermodynamic
quantities in the dual NRCQM theories in $d$ spatial dimensions at finite temperature and finite chemical potential.
Although we have not obtained the explicit uplifting of these geometries, neither do we know whether they
effectively can be
lifted to ten or eleven dimension, we discuss about the string theory and M-theory embedding
of the geometries which correspond to their asymptotic limits, i.e. AdS$_4$ and AdS$_6$ spaces, with
the aim that it can be useful for further investigations.

We are particularly interested in the DLCQ of conformal quantum field theories with plane wave boundary conditions
and their gravity dual description. These NRCQM theories are defined on plane wave backgrounds
in diverse dimensions.
We know that the plane wave metric is conformal to flat space.
The boundary plane wave structure can be explicitly shown by slicing the AdS metric, so for instance
one can start from
$\mathbb{R} \times S^{d+1}$ and take the Penrose limit for a particle moving with
large angular momentum along certain angular direction on the sphere. Let us start from
the $d+3$ dimensional AdS spacetime metric in global coordinates
\be
ds_{d+3}^2 = - (1+r^2) \,  dt^2 + \frac{1}{(1+r^2)} \, dr^2 + r^2 \, (d\theta^2 +
               \cos^2\theta \, d\varphi^2 + \sin^2\theta \, d\Omega_{d-1}^2) \, ,
\ee
where $d\Omega_{d-1}$ measures lengths on $S^{d-1}$ in $S^{d+1}$. So,
the boundary of this metric is $\mathbb{R} \times S^{d+1}$. Now, we can take the Penrose
limit by doing a scaling of the coordinates as follows
\be
x^+=t \, , \,\,\,\,\,\,\,\,
\frac{x^-}{2 R^2} = t-\varphi \, , \,\,\,\,\,\,\,\,
\theta = \frac{\tilde \rho}{R} \, , \,\,\,\,\,\,\,\,
r=R \, \tilde y \, ,
\ee
and then taking the limit $R\rightarrow \infty$,
while keeping $x^\pm$,$\tilde y$ and $\tilde \rho$ fixed, the metric becomes
\be
ds_{pp-wave}^2 = - (dx^+)^2 + {\tilde y}^2 \,
                   (- dx^+ \, dx^- - {\tilde\rho}^2 \, (dx^+)^2 + d{\tilde\rho}^2
                    + {\tilde\rho}^2 \, d\Omega_{d-1}^2) + \frac{{d\tilde y}^2}{{\tilde y}^2}
                    \, . \label{metric0}
\ee
After a further change of coordinates it can be brought into a form where the
slicing becomes explicit.

Now, if we consider the periodic identification of the
coordinate $x^-\sim x^- +2 \pi r^-$ the metric (\ref{metric0}) has a null direction, so
we cannot trust it. In order to overcome this issue we can
inject $N$ units of momentum along the $x^-$ direction. This prevents the circle to
collapse, so that we can trust the metric for gauge/gravity duality purposes.
Also we want to consider systems at finite temperature. We can do it
by introducing a black hole which asymptotically approaches the metric (\ref{metric0}).
Specifically we can begin with a Kerr-AdS$_{d+3}$ black hole, and take the same limit as
above.

Concerning the field theory side, the DLCQ of a general field theory on a
plane wave background leads to a quantum mechanical system with
particles in a harmonic oscillator potential, so that the Hamiltonian of the system reads
\be
-p_+ = \frac{{\vec p}_d^2}{(-4 p_-)} + (-p_-) \, {\vec x}_d^2 \, .
\ee
The isometries of the plane wave are reflected on the field theory. In addition,
if the field theory is conformal, the non-relativistic theory
will also be conformal. Let us consider this in more detail. To begin with, it is convenient to
introduce the Schr\"odinger group. The Schr\"odinger algebra results from the extension of the Galilean algebra
to the non-relativistic conformal group. The Galilean algebra in $d$ spatial
dimensions has the Hamiltonian $H$, momenta $P_i$, rotation generators $M_{ij}$,
Galilean boosts $K_i$ and the mass operator $M$ as generators, which satisfy certain commutation
relations. It can be embedded in the Poincar\'e algebra in $d+1$ spatial dimensions
and one time dimension, which has the momenta and rotation generators $\tilde P_i$
and $\tilde M_{ij}$, respectively, and their well-known commutation relations.
Now, by identifying $M=-\tilde P_-$, $H=-\tilde P_+$, $P_i=\tilde P_i$,
$M_{ij}=\tilde M_{ij}$ and $K_i=\tilde M_{-i}$, where the indices $+$ and $-$ denote
light cone coordinates $x^\pm = x^0 \pm x^{d+1}$, the embedding of the
Galilean algebra into the Poincar\'e algebra is explicitly done. If the dilatation generator $D$ is
added to the Galilean algebra there appears a number of commutation relations with
a constant known as the dynamical exponent. There is a special case when the dynamical exponent
takes value 2 where the algebra admits another extension,
by adding the special conformal transformations generator $C$.
This leads to the Schr\"odinger algebra. In addition, $H$, $D$ and $C$ generate a $SL(2,\mathbb{R})$
subgroup. These generators can be combined to give the raising/lowering operators
$L_\pm=\frac{1}{2}(H - C \mp i D)$ and the oscillator Hamiltonian
$H_{osc}=L_0=\frac{1}{2}(H + C)$ \cite{de Alfaro:1976je}. This is the Hamiltonian of the
NRCQM theory on the plane wave background.

Also, since there should be a certain string theory or M-theory description of the
gravity dual system, one would expect the isometries of the internal manifold
to play some role in the description of the symmetries of the NRCQM theory.
As mentioned, if one starts from a relativistic field theory which has a gravity dual description
it is expected that, after the DLCQ is done on both sides, the non-relativistic quantum
mechanical theory will also have a dual gravity description. The DLCQ limit of the gravity dual
is given by the identification of $x^-$ in the bulk. However, as we mentioned this identification
must be done avoiding the size of circle to vanish. This is achieved by injecting a large amount of momentum
$N\sim -p_- \, r^-$ along  $x^-$. This determines the region of parameters of the quantum mechanical theory
that should be tractable in terms of its gravity dual.

The paper is organized as follows. In section 2 we review some of the results of
\cite{Maldacena:2008wh} which are relevant to our calculations and studies.
In section 3 we calculate the DLCQ of the single parameter Kerr-AdS$_4$ black hole spacetime metric,
also carrying out the calculation of thermodynamic properties. Section 4 is devoted to
general higher dimensional single parameter Kerr-AdS$_{d+3}$ black hole metrics and their dual field
theories. After a brief general discussion for even and odd dimensional cases, we specialize the calculations
to the single parameter Kerr-AdS$_6$ and the
single parameter Kerr-AdS$_7$ black hole spacetime metrics. Additional general formulas are
introduced in the Appendix. Section 5 deals with the DLCQ of an alternative BTZ metric.
Also we discuss the embedding of the DLCQ of the
Kerr-AdS spaces in string theory and M-theory backgrounds. We explicitly
calculate the expressions of
thermodynamic properties of the dual NRCQM theories at finite temperature and finite
chemical potential on plane wave backgrounds of diverse dimensions. On the other hand, we have not
worked out the dual NRCQM theories explicitly, this is a very interesting and
difficult problem to deal with. In section 6 we discuss the results and conclusions.

\section{A review of Kerr-AdS$_5$ black holes and their DLCQ}

We begin with a brief review of a case studied by MMT. Consider the one-parameter
five-dimensional Kerr-AdS black hole metric written in the form
given by Gibbons et al \cite{Gibbons:2004ai}
\bea
ds^2_5 &=& -\frac{\Delta_r}{\rho^2} \left(dt - \frac{a}{\Xi} \cos^2\theta d\phi \right)^2 +
           \frac{\rho^2}{\Delta_r} dr^2 + r^2 \sin^2\theta d\psi^2  \nonumber \\
       & & + \frac{\rho^2}{\Delta_\theta} d\theta^2 + \frac{\Delta_\theta \cos^2\theta}{\rho^2}
           \left(a dt - \frac{r^2+a^2}{\Xi} d\phi \right)^2  \, , \label{metricKerrAdS5}
\eea
where we have used the following definitions
\bea
&& \Delta_r = (r^2+a^2)(1+r^2) - 2 m \, , \,\,\,\,\,\,\,\,\,\,\,\,
\Delta_\theta = 1-a^2 \sin^2\theta \, , \nonumber \\
&& \Xi = 1-a^2  \, ,   \,\,\,\,\,\,\,\,\,\,\,\,\,\,\,\,\,\,\,\,\,\,\,\,\,\,
\,\,\,\,\,\,\,\,\,\,\,\,\,\,\,\,\,\,\,\,\,\,\,\,\,\,\,\,\,\,\,\,
\rho^2 = r^2+a^2 \sin^2\theta \, .
\eea
Using the notation given in the introduction for this case we have $d=2$.
This metric was first derived by Hawking et al \cite{Hawking:1998kw}. One sets their parameter $l=1$ and
shifts $\theta\rightarrow\pi/2-\theta$. It is an Einstein metric satisfying $R_{\mu\nu}=-4 g_{\mu\nu}$.
Using the notation of Gibbons et al, the outer event horizon is located at the largest positive root
$r=r_+$ of $\Delta_r=0$. Its area is $A=2\pi^2 (r_+^2+a^2) r_+/\Xi$. The surface gravity $\kappa$ is
given by
\be
\kappa= r_+ (1+r_+^2) \left(\frac{1}{r_+^2+a^2}+\frac{1}{r_+^2} \right) - \frac{1}{r_+} \, .
\ee
The Hawking temperature is
\be
T=\frac{1}{\beta}=\frac{\kappa}{2\pi} \, .
\ee
The angular velocity relative to a non-rotating frame at infinity is
\be
\Omega= \frac{a(1+r_+^2)}{r_+^2+a^2} \, .
\ee
The action related to the five-dimensional Kerr-AdS black hole system above was obtained by
Hawking et al and it reads
\be
I_5 = \frac{\pi \beta}{4 \Xi} [m-(r_+^2+a^2)r_+^2] \, .
\ee
The expressions for the energy and angular momentum are given by \cite{Gibbons:2004ai}
\bea
E=\frac{\pi m (3-a^2)}{4(1-a^2)^2} \, , \,\,\,\,\,\,\,\,\,\,\, J=\frac{\pi m a}{2(1-a^2)^2} \, ,
\eea
where factors $R_{AdS_5}$ and $G_N^5$ have been dropped.
$J$ is the angular momentum $J_\phi$. For this configuration there is no
angular momentum along $\psi$.
The first law of thermodynamics reads as $\delta E=T \,\, \delta S+\Omega \,\, \delta J$, which by using
the above equations for $T$, $\Omega$ and $J$, is an exact differential provided that the entropy is
\be
S=\frac{A}{4} \, .
\ee
Thus, it can be integrated resulting the above expression for $E$, with the condition that $E=0$ for
$m=0$ for pure AdS$_5$ space.

The metric of Eq.(\ref{metricKerrAdS5}) is asymptotically AdS$_5$, however in these coordinates it is a
rotating Einstein universe. In order to get the $\mathbb{R} \times S^3$ boundary we have to change coordinates
as \cite{Hawking:1998kw}
\be
(1-a^2) {\hat r}^2 \cos^2\hat\theta=(r^2+a^2) \cos^2\theta \, , \,\,\,\,\,\,\,
{\hat r}^2 \sin^2\hat\theta=r^2 \sin^2\theta \, , \,\,\,\,\,\,\,
\hat\phi=\phi+a t \, ,
\ee
then the metric (\ref{metricKerrAdS5}) gets the asymptotic form \cite{Maldacena:2008wh}
\bea
d{\hat s}^2_5&=& -(1+{\hat r}^2) dt^2 + \frac{1}{1+{\hat r}^2-2 m/\Delta_{\hat\theta}} d{\hat r}^2
                +{\hat r}^2 (d{\hat \theta}^2+\cos^2{\hat \theta} d{\hat \phi}^2+
                \sin^2{\hat \theta} d\psi^2) \nonumber \\
&&              + \frac{2 m}{{\hat r}^2(1-a^2 \sin^2{\hat \theta}^2)^3} (dt-a \cos^2{\hat \theta}
                d{\hat \phi}^2)^2 + \cdot \cdot \cdot \, .
\eea

Now, we carry out the DLCQ of the metric of Eq.(\ref{metricKerrAdS5}). We use the following scaling
\bea
t &\equiv& x^+ \, , \label{t} \\
\phi &\equiv& \frac{1}{2 R^2} \left(-x^-+\frac{x^+}{\lambda}\right) \, , \label{phi} \\
\frac{1}{2 R^2} &\equiv& \lambda (1-a) \label{a} \, ,
\eea
where $\lambda$ is a parameter which takes into account the amount of momentum injected in the compact direction.
We should keep in mind that we consider the limit for $R\rightarrow \infty$ that leads to $a\rightarrow 1$.
Thus
\be
\lim_{R\rightarrow\infty}\Delta_r = (r^2+1)^2 - 2 m \, , \,\,\,\,\,\,\,\,
\lim_{R\rightarrow\infty}\Delta_\theta = \cos^2\theta \, , \,\,\,\,\,\,\,\,
\lim_{R\rightarrow\infty}\Xi = \frac{1}{R^2 \lambda} \, , \,\,\,\,\,\,\,\,
\lim_{R\rightarrow\infty}\rho^2 = r^2+\sin^2\theta \, . \nonumber
\ee
In this limit the metric (\ref{metricKerrAdS5}) becomes
\bea
ds^2_{5-DLCQ} &=& \frac{(r^2+\sin^2\theta)}{(r^2+1)^2 - 2 m} dr^2 - (1+r^2\sin^2\theta)(dx^+)^2
                  -\lambda (1+r^2) \cos^2\theta \,dx^+ \,dx^- + \nonumber \\
              & &  \frac{(r^2+\sin^2\theta)}{\cos^2\theta} d\theta^2 + r^2 \sin^2\theta d\psi^2
                  + m \frac{(-2 dx^+ + (dx^+-\lambda dx^-) \cos^2\theta)^2}{2(r^2+\sin^2\theta)} \, , \label{metricKerrAdS5DLCQ}
\eea
which was obtained by MMT.

The Hamiltonian $H$ and $N$ are calculated as follows
\bea
&& H = -P_+ = \frac{r^-}{2 R^2} \, (E-J) = \frac{1}{4} \, m \, \pi \, r^- \, \lambda \, , \label{H5} \\
&& N = -P_- \, r^- = \frac{(r^-)^2}{(2 R^2)^2} \, J = \frac{1}{8} \, m \, \pi \, (r^-)^2 \, \lambda^2 \label{N5} \, .
\eea
These expressions have been obtained in the $a\rightarrow 1$ limit.  Using the
condition of vanishing $\Delta_r(r_+)=0$ at $r_+$, it leads to
$m=(r_+^2+a^2)(r_+^2+1)/2$. Therefore, the expressions (\ref{H5}) and (\ref{N5}) are the ones of MMT.
In this limit the temperature, the chemical potential and the entropy of the system are respectively
\bea
&& T = \frac{r_+}{\pi} \, , \label{T5} \\
&& \mu = \frac{1}{r^- \lambda} \, \frac{(r_+^2-1)}{(r_+^2+1)} \, , \label{mu5} \\
&& S = \frac{1}{4} \, \pi^2 \, \lambda \, r_+ \, (1+r_+^2) \, r^- \, . \label{S5}
\eea
In addition, the five-dimensional action in this limit becomes
\be
I_5 = \frac{\pi^2 R^2 \lambda (1-r_+^4)}{8 r_+}  \, ,
\ee
which vanishes for $r_+=1$.
We can also explicitly check that the first law of thermodynamics $\delta H=T \delta S-\mu \delta N$
is satisfied using the expressions (\ref{H5}-\ref{S5}). Also MMT have obtained the radius of the circle along
$x^-$ at the horizon in units of the Planck length, which is given by
\be
R^-_{physical}=r^- \, \lambda \, \frac{(r_+^2+1) \, \cos^2\theta}{2 \sqrt{r_+^2+\sin^2\theta}} \, R_{AdS_5} \, .
\ee
This radius shrinks at $\theta=\pi/2$ at any value of $r$.

Now, if one changes coordinates as
\be
y^2 = \lambda \, (r^2+1) \, \cos^2\theta \, , \,\,\,\,\,\,\,\,
\rho^2 \, y^2 = r^2 \, \sin^2\theta \, , \label{newcoord}
\ee
it can be shown that the metric (\ref{metricKerrAdS5DLCQ}) becomes \cite{Maldacena:2008wh}
\bea
d{\tilde s}^2_{5-DLCQ} &=& \left(1-\frac{2 \, m}{(1+\lambda \, \rho^2)^2} \, \frac{\lambda^2}{y^4} \right)^{-1} \,
                  \frac{dy^2}{y^2}
                  + y^2 (-dx^+ \, dx^- - \rho^2 \, (dx^+)^2 + d\rho^2 + \rho^2 \, d\psi^2) \nonumber \\
              & & - (dx^+)^2 + \frac{m \, \lambda}{y^2}
              \frac{((1+2\lambda\rho^2)dx^++\lambda dx^-)^2}{2(1+\lambda \rho^2)^3} + \cdot \cdot \cdot \, ,
                  \label{metricKerrAdS5DLCQrhoy}
\eea
which has a similar asymptotic structure as the metric (\ref{metric0}).
The dots indicate terms of order $y^{-4}$, as well as deformations of the plane wave
boundary metric like $+[1/(y^4 \rho^6)+3/(y^2 \rho^4)+3/\rho^2] \, d\rho^2$, and deformations
arising from the last term of the metric (\ref{metricKerrAdS5DLCQ}) which are proportional to $m$.
Thus, the region where the size of the circle shrinks corresponds to the limit $y\rightarrow 0$ and $\rho\rightarrow \infty$. However, there are no large contributions to the above described thermodynamic properties
from this region. In this way we assume that this region of the metric can be ignored. The
explicit NRCQM theory dual to this system is not known, but it is defined in 2 spatial dimensions at
finite temperature and finite chemical potential.

In what follows we extend this analysis to Kerr-AdS black hole backgrounds of diverse number of dimensions.
All the backgrounds considered in the following sections correspond to AdS$_{d+3}$ spaces with a plane wave
boundary.

\section{DLCQ of the Kerr-AdS$_4$ black hole}

Let us consider now the case when $d=1$. The NRCQM theory is defined on one spatial dimension.
The Kerr-AdS$_4$ black hole metric was obtained by Carter \cite{Carter:1968ks}. We use the notation of Gibbons et al,
\cite{Gibbons:2004ai}, however we take $l=1$ and change $\theta\rightarrow \pi/2 - \theta$. Thus
\be
ds^2_4 = -\frac{\Delta_r}{\rho^2} \left(dt - \frac{a}{\Xi} \cos^2\theta d\phi \right)^2 +
           \frac{\rho^2}{\Delta_r} dr^2  +
        \frac{\rho^2}{\Delta_\theta} d\theta^2 + \frac{\Delta_\theta \cos^2\theta}{\rho^2}
           \left(a dt - \frac{r^2+a^2}{\Xi} d\phi \right)^2  \, , \label{metricKerrAdS4}
\ee
where we have used the following definitions
\bea
&& \Delta_r = (r^2+a^2)(1+r^2) - 2 m \, r \, , \,\,\,\,\,\,\,\,\,\,\,\,
\Delta_\theta = 1-a^2 \sin^2\theta \, , \nonumber \\
&& \Xi = 1-a^2  \, , \,\,\,\,\,\,\,\,\,\,\,\,\,\,\,\,\,\,\,\,\,\,\,\,\,\,\,\,\,
\,\,\,\,\,\,\,\,\,\,\,\,\,\,\,\,\,\,\,\,\,\,\,\,\,\,\,\,\,\,\,\,
\rho^2 = r^2+a^2 \sin^2\theta \, .
\eea
In this case $R_{\mu\nu}=-3 g_{\mu\nu}$. The outer horizon is located at the largest positive root
$r=r_+$ of $\Delta_r=0$. The area of the event horizon is $A=4 \pi (r_+^2+a^2)/\Xi$. The surface gravity
$\kappa$ is given by
\be
\kappa= r_+ \frac{(1+a^2+3r_+^2-a^2 r_+^{-2})}{2 (r_+^2+a^2)} \, .
\ee
The Hawking temperature is
\be
T=\frac{1}{\beta}=\frac{\kappa}{2\pi} \, .
\ee
The angular velocity relative to a non-rotating frame at infinity is
\be
\Omega= \frac{a(1+r_+^2)}{r_+^2+a^2} \, .
\ee
The action related to the four-dimensional Kerr-AdS black hole is given by
\be
I_4 = -\frac{\pi (r_+^2+a^2)^2 (r_+^2-1)}{\Xi(3r_+^4+(1+a^2)r_+^2-a^2)} \, . \label{action4}
\ee
The energy and angular momentum are \cite{Gibbons:2004ai}
\bea
E=\frac{m}{(1-a^2)^2} \, , \,\,\,\,\,\,\,\,\,\,\, J=\frac{m a}{(1-a^2)^2} \, .
\eea
With these expressions for $E$, $T$, $\Omega$ and $J$ it is straightforward to verify that
the first law of thermodynamics $\delta E = T \,\, \delta S + \Omega \,\, \delta J$
is satisfied, provided that the entropy is $S=\frac{A}{4}$.

Since we are interested in the discrete light cone quantization of the Kerr-AdS$_4$ black hole
background, the idea is to perform a scaling on the metric (\ref{metricKerrAdS4}) and on the
thermodynamic properties as well, in order to be able to obtain explicit expressions for the
Hamiltonian, the particle number, entropy, chemical potential and temperature
of the dual non-relativistic conformal quantum mechanical theory. The scaling is the same as
in the previous case, given by Eqs.(\ref{t}-\ref{a}).
As in the five-dimensional case $\lambda$ is a parameter
which takes into account the amount of momentum injected in the compact direction $x^-$.
We consider the limit for $R\rightarrow \infty$ so
that $a\rightarrow 1$. Thus, we have the following expressions for the limits
\be
\lim_{R\rightarrow\infty}\Delta_r = (r^2+1)^2 - 2 m r \, , \,\,\,\,\,\,\,
\lim_{R\rightarrow\infty}\Delta_\theta = \cos^2\theta \, , \,\,\,\,\,\,\,
\lim_{R\rightarrow\infty}\Xi = \frac{1}{R^2 \lambda} \, , \,\,\,\,\,\,\,
\lim_{R\rightarrow\infty}\rho^2 = r^2+\sin^2\theta \, .
\ee
In this limit the metric Eq.(\ref{metricKerrAdS4}) becomes
\bea
ds^2_{4-DLCQ} &=& \frac{(r^2+\sin^2\theta)}{(r^2+1)^2 - 2 m r} dr^2 - (1+r^2\sin^2\theta)(dx^+)^2
                  -\lambda (1+r^2) \cos^2\theta \,dx^+ \,dx^-  \nonumber \\
              & & + \frac{(r^2+\sin^2\theta)}{\cos^2\theta} d\theta^2
                  + m \, r \frac{(-2 dx^+ + (dx^+-\lambda dx^-) \cos^2\theta)^2}{2(r^2+\sin^2\theta)} \, . \label{metricKerrAdS4DLCQ}
\eea
Now, we calculate the Hamiltonian $H$ and $N$ and obtain
\bea
&& H = -P_+ = \frac{r^-}{2 R^2} \, (E-J) = \frac{1}{4} \, m  \, r^- \, \lambda \, , \label{H4} \\
&& N = -P_- \, r^- = \frac{(r^-)^2}{(2 R^2)^2} \, J = \frac{1}{4} \, m \, (r^-)^2 \, \lambda^2 \label{N4} \, .
\eea
These expressions have been obtained in the $a\rightarrow 1$ limit. $m$ is set using the
condition of vanishing $\Delta_r(r_+)=0$ at the outer black hole horizon, which leads to
$m=(r_+^2+a^2)(r_+^2+1)/(2 r_+)$.
In this limit the temperature, the chemical potential and the entropy of the system are respectively
\bea
&& T = \frac{3 r_+^2 - 1}{4 \pi r_+} \, , \label{T4} \\
&& \mu = \frac{1}{r^- \lambda} \, \frac{(r_+^2-1)}{(r_+^2+1)} \, , \label{mu4} \\
&& S = \frac{1}{2} \, \pi \, \lambda \, (1+r_+^2) \, r^- \, . \label{S4}
\eea
We have explicitly checked that the first law of thermodynamics $\delta H=T \,\, \delta S-\mu \,\, \delta N$
is satisfied using the expressions (\ref{H4}-\ref{S4}). The radius of the circle along
$x^-$ at the horizon in units of the Planck length is
\be
R^-_{physical}=r^- \, \lambda \, \frac{(r_+^2+1) \, \cos^2\theta}{2 \sqrt{r_+^2+\sin^2\theta}} \, R_{AdS_4} \, .
\ee
This radius shrinks at $\theta=\pi/2$ at any value of $r$.
However, the same discussion as in the Kerr-AdS$_5$ case applies.
Therefore, there are no large contributions to the above described thermodynamic properties
from this region of the metric, so that we assume that this region of the metric can be ignored.

Also, in the limit above the four dimensional action (\ref{action4}) becomes
\be
I_4 = -\frac{\pi (r_+^2+1)^2 (r_+^2-1) R^2 \lambda}{(3r_+^4+2r_+^2-1)} \, .
\ee

In addition, if one changes coordinates as Eq.(\ref{newcoord}) the metric (\ref{metricKerrAdS4DLCQ}) reads as
\bea
d{\tilde s}^2_{4-DLCQ} &=& \left(1-\frac{2 \, m}{(1+\lambda \, \rho^2)^{3/2}} \, \frac{\lambda^{3/2}}{y^{3}} \right)^{-1} \,
                  \frac{dy^2}{y^2}
                  + y^2 (-dx^+ \, dx^- - \rho^2 \, (dx^+)^2 + d\rho^2) \nonumber \\
              & & - (dx^+)^2 + \frac{m \, \lambda^{1/2}}{y}
              \frac{((1+2\lambda\rho^2)dx^++\lambda dx^-)^2}{2(1+\lambda \rho^2)^{5/2}} + \cdot \cdot \cdot \, ,
                  \label{metricKerrAdS5DLCQrhoy}
\eea
which has an asymptotic form similar to the metric (\ref{metric0}) with $d=1$. In this case dots
indicate terms of order $y^{-4}$, and deformations of the plane wave
boundary metric proportional to $d\rho^2$, and other deformations
which come from the last term of the metric (\ref{metricKerrAdS4DLCQ}), and are proportional to $m$.
We would expect the Hamiltonian of the NRCQM theory to be of the form
$H_{osc}$ in one spatial dimension. There should be two parameters in this theory. One should be $N$, corresponding
to the number of non-relativistic particles moving along $x^-$, and if the background uplifts to
eleven dimensional supergravity, there should be a second parameter counting the number of M2-branes whose
backreaction induces the eleven dimensional spacetime.

On the other hand, the M-theory maximally supersymmetric embedding of AdS$_4$ is
AdS$_4 \times S^7$. This configuration
is the near horizon limit of a large number of M2-branes, the isometries of this metric are $SO(2,3)$ and
$SO(8)$, and it preserves 16 supercharges. Using the radial coordinate $U=r/\alpha'$, where $\alpha'=l_s^2$,
it is possible to find a region where it is the dual gravity description of
a supersymmetric Yang Mills theory in the large $N_c$ limit.
This region is for $U<<g^2_{YM}$, where $g_{YM}$ is the coupling of the gauge theory. This supergravity
description corresponds to a conformal field theory with $SO(8)$ R-symmetry
\cite{Maldacena:1997re,Itzhaki:1998dd}. The isometries of the metric
become symmetries in the field theory, and they should also be somehow reflected in the
DLCQ version of it. In general, the uplifting of AdS$_4$ to eleven dimensional supergravity can have
different seven dimensional manifolds. For instance it is possible to
obtain AdS$_4 \times  \tilde{S}^7$ (squashed seven-sphere), also there can be AdS$_4 \times Q^{1,1,1}$
and AdS$_4 \times N^{0,1,0}$ solutions of eleven dimensional supergravity.
Their 2+1 dimensional dual SYM theories (at zero temperature and zero chemical potential) preserve ${\mathcal{N}}=1$, ${\mathcal{N}}=2$, ${\mathcal{N}}=3$ supersymmetries, respectively (see \cite{Gursoy:2002tx} and references therein).
The isometries of $\tilde{S}^7$, $Q^{1,1,1}$ and $N^{0,1,0}$ are
$SO(5) \times SO(3)$, $SU(2)^3 \times U(1)$ and $SU(3) \times SU(2)$, respectively,
and they should be reflected in the dual NRCQM theories at finite temperature and finite
chemical potential after the DLCQ is performed, starting from a black hole
instead of the AdS space.
If one considers a finite temperature version of the theory,
a possibility is that at low temperature the system should be described by
a thermal AdS$_4$ times some seven dimensional manifold, and as temperature increases
the system should undergo a Hawking-Page type of phase transition leading to the Kerr-AdS$_4$ black hole
embedded in eleven dimensions.

For example, at zero temperature and zero chemical potential
the CFT which is dual to the background AdS$_4 \times Q^{1,1,1}$
corresponds to the theory on M2-branes on the tip of a
cone on the seven-dimensional manifold \cite{Fabbri:1999ag,Dall'Agata:1999hh,Ahn:1999ec,Oh:1998qi,
Fabbri:1999ay,Acharya:1998db,Morrison:1998cs,Herzog:2000rz}.
The moduli space of vacua of this SYM theory is isomorphic to $Q^{1,1,1}$.
The theory has a Coulomb branch described by fields in the vector
multiplet and a Higgs branch described by fields in chiral multiplets.
The theory whose Higgs branch is dual to the conifold above has the following field content.
There are fundamental fields which are doublets with respect to the
flavour group $SU(2)^3$ : $A_i$, $B_i$, $C_i$ with $i = 1, \, 2,$ i.e.,
under the flavour group the fields transform as
$A_i = (2, 1, 1)$, $B_i = (1, 2, 1)$, $C_i = (1, 1, 2)$.
The gauge symmetry is $SU(N_c) \times SU(N_c) \times SU(N_c)$, with elementary
degrees of freedom transforming in the fundamental and anti-fundamental representations
of the $SU(N_c)$ groups: $A_i = (N_c, {\bar{N_c}}, 1)$, $B_i = (1,N_c,\bar{N_c})$, $C_i = (\bar{N_c}, 1, N_c)$.
These fields have conformal weight $c = 1/3$, hence one can construct gauge invariant
operators of the form $X^{ijk} = A^i B^j C^k$
out of them. These eight operators are singlets under the global symmetries and have
conformal weight equal to one. On the other hand, very recently Franco, Hanany, Park and
Rodr\'{\i}guez-G\'omez have investigated M2-brane theories for generic toric
singularities and studied the conifold over $Q^{1,1,1}$ \cite{Franco:2008um}.
Their proposal is a theory of the type worked out by Aharony, Bergman, Jafferis and Maldacena
\cite{Aharony:2008ug}.
In \cite{Franco:2008um}, the moduli space has been explicitly calculated, also
the symmetries have been studied and it has been shown that the theory is connected with other theories
through RG flows suggested by crystal models\footnote{We thank Sebasti\'an Franco for
calling our attention about the work of reference \cite{Franco:2008um}.}.

In the case of the AdS$_4 \times N^{0,1,0}$ space, the dual CFT has gauge group $SU(N_c) \times SU(N_c)$
and a flavor group $SU(3)$. There will be two hypermultiplets, $u_1$, $u_2$ and $v_1$, $v_2$
transforming in the $(3, N_c, \bar{N_c})$ and $(\bar{3}, \bar{N_c} , N_c)$
representations and two chiral multiplets, $Y_{(1)}$, $Y_{(2)}$ in the adjoint representation of
$SU(N_c)$. There is a superpotential of the form
\be
V \sim g_i \, Tr(Y_{(i)} \vec{u} \cdot \vec{v}) + \alpha_i \, Tr(Y_{(i)} Y_{(i)}) \, ,
\ee
where $g_i$ are the gauge couplings of each $SU(N_c)$ group and $\alpha_i$ are the Chern Simons coefficients.
Other properties of this theory, including the KK spectrum of the compactifications, as well as
different checks of the duality have been studied in \cite{Termonia:1999cs,Billo:2000zr}.

In addition, one can consider for instance the maximally supersymmetric plane
wave background of eleven-dimensional
supergravity, which can be obtained as a Penrose limit of  AdS$_7 \times S^4$ or
AdS$_4 \times S^7$ \cite{Blau:2002dy}.
Berenstein, Maldacena, and Nastase (BMN) \cite{Berenstein:2002jq} proposed a supersymmetric matrix model describing
the discrete light cone quantization of M-theory in this background. The corresponding
Hamiltonian can be thought of as a massive deformation of the BFSS matrix model \cite{Banks:1996vh}
$H = H_0 + H_\mu$, where $H_0$ is the BFSS Hamiltonian describing DLCQ of M-theory
in eleven flat dimensions, while $H_\mu$ is the massive deformation. Motl, Neitzke and
Sheikh-Jabbari \cite{Motl:2003rw} carried out a suitable projection in the BMN matrix model, and they added extra
fermionic degrees of freedom (0-8 strings). These fermionic degrees of freedom
are introduced to account for the coupling of
the D0-branes to the gauge theory on the boundary. Also it can be though of as they are necessary to cancel the anomaly in the open membrane worldvolume theory \cite{Horava:1995qa}. The result is a DLCQ description of M-theory
on the orbifolded plane wave which is the Penrose limit of AdS$_7 \times S^4/\mathbb{Z}_2$ and
AdS$_4/\mathbb{Z}_2 \times S^7$.

\section{DLCQ of higher dimensional Kerr-AdS black holes}

We consider the single parameter Kerr-AdS black hole metrics in $d+3$ dimensions obtained by
Hawking, Hunter and Taylor-Robinson \cite{Hawking:1998kw}. As for the case of the stationary asymptotically
flat solutions of Myers and Perry \cite{Myers:1986un}, the single rotation parameter Kerr-AdS
black hole in higher dimensions of Hawking et al follows from the four dimensional system.
We use $d$ to denote the number of spatial dimensions of the dual NRCQM theory. Again,
we use the notation given in \cite{Hawking:1998kw}, however we take $l=1$ and change
$\theta\rightarrow \pi/2 - \theta$. Thus
\bea
ds^2_{d+3} &=& -\frac{\Delta_r}{\rho^2} \left(dt - \frac{a}{\Xi} \cos^2\theta d\phi \right)^2 +
           \frac{\rho^2}{\Delta_r} dr^2  +
           \frac{\rho^2}{\Delta_\theta} d\theta^2 + \frac{\Delta_\theta \cos^2\theta}{\rho^2}
           \left(a dt - \frac{r^2+a^2}{\Xi} d\phi \right)^2  \nonumber \\
       & &  + r^2 \sin^2\theta d\Omega^2_{d-1} \, , \label{metricKerrAdSd}
\eea
where we have used the definitions
\bea
&& \Delta_r = (r^2+a^2)(1+r^2) - 2 m \, r^{2-d} \, , \,\,\,\,\,\,\,\,\,\,\,\,
\Delta_\theta = 1-a^2 \sin^2\theta \, , \nonumber \\
&&\Xi = 1-a^2  \, , \,\,\,\,\,\,\,\,\,\,\,\,\,\,\,\,\,\,\,\,\,\,\,\,\,\,\,\,\,
\,\,\,\,\,\,\,\,\,\,\,\,\,\,\,\,\,\,\,\,\,\,\,\,\,\,\,\,\,\,\,\,\,\,\,\,
\rho^2 = r^2+a^2 \sin^2\theta \, . \nonumber
\eea
Now, we perform the DLCQ of the metric of Eq.(\ref{metricKerrAdSd}). First we use the scaling of
Eqs.(\ref{t}-\ref{a}).
We therefore have the parameter $\lambda$ as in the previous cases.
The limit $R\rightarrow \infty$ is equivalent to  $a\rightarrow 1$. We
also consider the limits of the following expressions
\be
\lim_{R\rightarrow\infty}\Delta_r = (r^2+1)^2 - 2 m r^{2-d} \, , \,\,\,\,\,\,\,
\lim_{R\rightarrow\infty}\Delta_\theta = \cos^2\theta \, , \,\,\,\,\,\,\,
\lim_{R\rightarrow\infty}\Xi = \frac{1}{R^2 \lambda} \, , \,\,\,\,\,\,\,
\lim_{R\rightarrow\infty}\rho^2 = r^2+\sin^2\theta \, . \,\,\,\,\,\,\,
\ee
In this limit the metric Eq.(\ref{metricKerrAdSd}) becomes
\bea
ds^2_{d+3-DLCQ} &=& \frac{(r^2+\sin^2\theta)}{(r^2+1)^2 - 2 m r^{2-d}} dr^2 - (1+r^2\sin^2\theta)(dx^+)^2
                  -\lambda (1+r^2) \cos^2\theta \,dx^+ \,dx^-  \nonumber \\
              & & + \frac{(r^2+\sin^2\theta)}{\cos^2\theta} d\theta^2
                  + m \, r^{2-d} \frac{(-2 dx^+ + (dx^+-\lambda dx^-) \cos^2\theta)^2}{2(r^2+\sin^2\theta)}
                  \nonumber \\
              & &  + r^2 \sin^2\theta d\Omega^2_{d-1}  \, . \label{metricKerrAdSdDLCQ}
\eea
The expressions for the energy, angular momentum, angular velocity, temperature and entropy are given in
the Appendix. Now, we consider the particular cases of Kerr-AdS$_6$ and Kerr-AdS$_7$ black holes, and perform
their discrete light cone quantization. General expressions for the Hamiltonian, the chemical potential,
the temperature and the entropy of Kerr-AdS$_{d+3}$ black holes for odd and even dimensions are described
in the Appendix. Also in the Appendix we write down the general expression to display the asymptotic
structure of the AdS-plane wave metric arising from Eq.(\ref{metricKerrAdSdDLCQ}).

\subsection{DLCQ of Kerr-AdS$_6$ black hole}

The case when $d=3$ is particularly interesting because it is relevant for
NRCQM theories defined in 3 spatial dimensions at finite temperature and finite chemical potential.
The area of the event horizon is $A=8 \pi^2 r_+^2 (r_+^2+a^2)/(3 \, \Xi)$. The surface gravity
$\kappa$ is given by
\be
\kappa= r_+ (1+r_+^{2})\left(\frac{1}{(r_+^2+a^2)}+\frac{1}{r_+^2}\right)-\frac{1-r_+^2}{2 r_+} \, .
\ee
The Hawking temperature, $T$, has the same definition as in the preceding sections.
The angular velocity relative to a non-rotating frame at infinity is
\be
\Omega= \frac{a(1+r_+^2)}{r_+^2+a^2} \, .
\ee
The energy and angular momentum are \cite{Gibbons:2004ai}
\bea
E=\frac{2 \pi m}{3(1-a^2)} \left(\frac{1}{(1-a^2)}+1\right) \, ,
\,\,\,\,\,\,\,\,\,\,\, J=\frac{2 \pi m a}{3(1-a^2)^2} \, .
\eea
The angular momentum corresponds to the direction $\phi$. Rotations in other planes are
not considered since the only non-vanishing parameter associated to rotations is $a$, and we
set to zero any other rotation parameter.

The entropy is $S=\frac{A}{4}$, so that with the expressions above for the energy, temperature,
angular momentum and angular velocity, the first law of thermodynamics
$\delta E = T \,\, \delta S + \Omega \,\, \delta J$ is an exact differential.
After we perform the DLCQ of the metric of Eq.(\ref{metricKerrAdSd}) we get
Eq.(\ref{metricKerrAdSdDLCQ}), and particularly in the present case for $d+3=6$ the metric becomes
\bea
ds^2_{6-DLCQ} &=& \frac{(r^2+\sin^2\theta)}{(r^2+1)^2 - \frac{2 m}{r}} dr^2 - (1+r^2\sin^2\theta)(dx^+)^2
                  -\lambda (1+r^2) \cos^2\theta \,dx^+ \,dx^-  \nonumber \\
              & & + \frac{(r^2+\sin^2\theta)}{\cos^2\theta} d\theta^2
                  + \frac{m}{r} \frac{(-2 dx^+ + (dx^+-\lambda dx^-) \cos^2\theta)^2}{2(r^2+\sin^2\theta)}
                  \nonumber \\
              & &  + r^2 \sin^2\theta d\Omega^2_{2}  \, . \label{metricKerrAdS6DLCQ}
\eea
Now, we calculate the Hamiltonian $H$ and number of particles $N$, obtaining
\bea
&& H = -P_+ = \frac{r^-}{2 R^2} \, (E-J) = \frac{1}{2} \, \pi \, m  \, r^- \, \lambda \, , \label{H6} \\
&& N = -P_- \, r^- = \frac{(r^-)^2}{(2 R^2)^2} \, J = \frac{1}{6} \, \pi \, m \, (r^-)^2 \, \lambda^2 \label{N6} \, .
\eea
These expressions have been obtained in the $a\rightarrow 1$ limit. The parameter $m$ is set using the
condition of vanishing $\Delta_r(r_+)=0$ at the outer horizon. This leads to $m=(r_+^2+a^2)(r_+^2+1)r_+/2$.
In this limit the temperature, the chemical potential and the entropy of the system are
\bea
&& T = \frac{5 r_+^2 + 1}{4 \pi r_+} \, , \label{T6} \\
&& \mu = \frac{1}{r^- \lambda} \, \frac{(r_+^2-1)}{(r_+^2+1)} \, , \label{mu6} \\
&& S = \frac{1}{3} \, \pi^2 \, \lambda \, r_+^2 \, (1+r_+^2) \, r^- \, , \label{S6}
\eea
respectively.
We have explicitly checked that the first law of thermodynamics $\delta H=T \,\, \delta S-\mu \,\, \delta N$
is satisfied using the expressions (\ref{H6}-\ref{S6}). The radius of the circle along
$x^-$ at the horizon in units of the Planck length is
\be
R^-_{physical}=r^- \, \lambda \, \frac{(r_+^2+1) \, \cos^2\theta}{2 \sqrt{r_+^2+\sin^2\theta}} \, R_{AdS_6} \, .
\ee
This radius shrinks at $\theta=\pi/2$ at any value of $r$.
However, the same discussion as in the Kerr-AdS$_4$ and Kerr-AdS$_5$ black hole systems applies.
Therefore, we conclude that there are no large contributions to the above described thermodynamic properties
from this region of the metric, so that we assume that this region of the metric can be ignored.
Also it is possible to obtain an explicit plane wave form which asymptotes the metric (\ref{metric0}).
The resulting metric is given by setting $d=3$ in the metric (\ref{metricKerrAdSdDLCQrhoy}) given in the Appendix.

In the present case, asymptotically the Kerr-AdS black hole metric is AdS$_6$.
The AdS$_6$ space can be uplifted to a ten dimensional solution of massive type IIA
supergravity. In this case the internal manifold is a four-sphere. Indeed, the ten dimensional uplifting is not
a direct product but a fibration of the form
AdS$_6$$\otimes S^4$, which turns out to be a solution of Romans' massive type IIA supergravity
\cite{Romans:1985tz}. It has the $SO(2,5) \times SO(4)$ isometry groups. This solution was obtained
by Brandhuber and Oz \cite{Brandhuber:1999np} and it corresponds to the near horizon limit of the D4-D8-brane system
with $N_c$ D4-branes and 16 D8-branes.
This supergravity solution is dual to the large $N_c$ limit of a family of five dimensional conformal field theories
known as Seiberg fixed points which have a parameter $N_f$ related to a global symmetry $E_{N_f+1}$.
More specifically, the solution describes the dynamics of a large number of D4-branes parallel to D8-branes located at the orbifold fixed planes.

It is very interesting to understand how this configuration arises in string theory.
Let us consider $N_c$ coincident D5-branes in type I string theory on the ${\mathbb{R}}^9 \times S^1$ background.
The D5-branes are wrapping the circle. If one does a T-duality on the circle, the
configuration results in type I' theory compactified
on the interval $S^1/\mathbb{Z}_2$ with two orientifolds (O8 planes) located at the fixed points.
The D5-branes become D4-branes and there are 16 D8-branes located at points on the
interval. They cancel the -16 units of D8-brane charge carried by the two O8 planes, and the
dynamics in the region between D8-branes is described by Romans' massive type IIA supergravity
\cite{Polchinski:1995mt,Polchinski:1995df}.

If we consider just one D4-brane worldvolume, the theory preserves ${\mathcal{N}}=2$
supersymmetries in five dimensions. Depending on $N_f$ the global symmetry groups are
$E_{N_f+1}$, with $E_8$, $E_7$, $E_6$, $E_5=Spin(10)$, $E_4=SU(5)$,
$E_3=SU(3)\times SU(2)$, $E_2=SU(2)\times U(1)$ and $E_1=SU(2)$
\cite{Seiberg:1996bd,Intriligator:1997pq}. We can also comment on the field content of the gauge theory.
The vector multiplet in five
dimensions has one real scalar component, a vector gauge field and a spinor, while the
hypermultiplet has four real scalars and a fermion. This theory has a Coulomb branch
when the real scalar has a VEV, and a Higgs branch when the scalar in the hypermultiplet
is excited. The hypermultiplet in the antisymmetric representation is massless.
In addition, the mass of the fundamental hypermultiplets is given by the relative
position of the D8-branes with respect to the D4-brane. Now, if we place $N_c$ parallel D4-branes on
top of each other the gauge group is $Sp(N_c)$ with one vector multiplet and hypermultiplets.
The scalar component of the vector multiplet describes the Coulomb branch $R^+$ of the gauge theory,
and the first components of the hypermultiplets describe the Higgs branch.
The global symmetries in the field theory are generated by the action of the
$SU(2)_R \times SU(2) \times SO(2 N_f ) \times U(1)$ groups and also by the conformal group $SO(2,5)$.
The first $SU(2)$ group corresponds to the R-symmetry,
the supercharges and the scalars in the hypermultiplets are doublets under the action of this group.
The second $SU(2)$ group is associated with the hypermultiplet
in the antisymmetric representation and the rest is associated with the hypermultiplets
in the fundamental representation and instantons. If the D4-brane is in the origin of the Coulomb
branch we have a fixed point on the gauge theory side,
i.e. a CFT, and the global symmetry is enhanced to $SU(2)\times E_{N_f+1}$.
If we had just one D5-brane in the initial type I string theory the dual gravity system
would be represented by $N_f$ D8-branes located on a $O8$ plane, (16 - $N_f$)
D8-branes in another fixed plane, and a D4-brane which can move between them. The
position of the D4-brane is parameterized by the scalar in the vector multiplet.
If the VEV of this scalar vanishes, this means that the D4-brane is on the $O8$ plane,
which leads to a gauge theory with $SU(2)$ R-symmetry group, and the theory has $N_f$
quarks. On the other hand, we can also give a non-zero VEV for the scalar in the vector
multiplet leading to a theory with a $U(1)$ symmetry with $N_f$ "electrons".

The ten dimensional metric is \cite{Brandhuber:1999np}
\be
ds^2 = \left(\frac{3}{4 \pi} C (8-N_f) \sin\alpha \right)^{-1/3} \,
       \left(N_c^{-1/2} U^2 dx_5^2 +  N_c^{1/2} \frac{9 dU^2}{4 U^2}
       + N_c^{1/2} d\Omega_4^2  \right) \, ,
\ee
where $C$ is an arbitrary parameter of the solution \cite{Polchinski:1995df}, and
\be
d\Omega_4^2 = d\alpha^2 + \cos^2\alpha \, d\Omega_3^2 \, ,
\ee
while the dilaton is
\be
e^\Phi = N_c^{-1/4} \, C \, \left(\frac{3}{4 \pi} C (8-N_f) \sin\alpha \right)^{-5/6} \, .
\ee
So we see explicitly a fibration of AdS$_6$ over $S^4$. This is the
most general form of a metric that has the isometry of an AdS$_6$ space. The space
has a boundary at $\alpha = 0$ which corresponds to the location of the O8 plane. The
boundary is of the form AdS$_6 \times S^3$. In addition to the $SO(2, 5)$ AdS$_6$ isometries, the
ten dimensional space has also $SO(4)$ isometries associated with the spherical part of the
metric above. In general $S^4$ has the $SO(5)$ isometry group. However, this is reduced due
to the warped product structure which leads to the isometry generated by the $SO(4)$ group.

If one considers the description at the boundary $\alpha = 0$,
the dilaton blows up and type I' theory becomes strongly coupled.
In the weakly coupled dual heterotic string description this is seen as an enhancement of
the gauge symmetry to $E_{N_f+1}$. One can see this enhancement of the gauge symmetry in
the type I' description by analyzing the D0-brane dynamics near the orientifold plane
\cite{Bergman:1997py,Matalliotakis:1997qe,Bachas:1997kn}.
This means that we have $E_{N_f+1}$ vector fields that propagate on the AdS$_6 \times S^3$
boundary, as in the Horava-Witten picture \cite{Horava:1996ma}. Also the scalar curvature of the background
blows up at the boundary. In the dual heterotic description the dilaton is small
but the curvature is large. For large $N_c$ there is a region, $\sin\alpha >> N_c^{-3/10}$, where both
curvature and dilaton are small and thus we can trust supergravity.

Other supergravity solutions with AdS spaces of lower dimensions were constructed in \cite{Nunez:2001pt}.
These solutions are dual to twisted field theories which are the worldvolume
theories of D4-branes wrapped on 2 and 3-cycles, as well as NS-fivebranes wrapped on 2-cycles.
All of these examples, including the AdS$_6$ case,
turn out to be spontaneous compactifications of massive type IIA supergravity
to Romans' $F(4)$ gauged supergravity \cite{Romans:1985tw}.

It is interesting to mention that Lowe, Nastase and Ramgoolam \cite{Lowe:2003qy}
proposed a Matrix theory approach to Romans'
massive type IIA supergravity. They applied the procedure of Matrix theory compactifications to the
proposal of the massive type IIA string theory as M-theory on a twisted torus developed by Hull \cite{Hull:1998vy}.
They obtained a Matrix theory which is a supersymmetric Yang Mills theory on a large number of D3-branes with a
space dependent non-commutativity parameter. They showed that energies of a class of physical excitations
of the supersymmetric Yang Mills theory have the correct symmetry expected from
massive type IIA string theory in a light cone quantization.

Although we are not able
to give a precise definition of the NRCQM theory in 3 spatial dimensions the physics described
in the paragraphs above should provide some hints. As we have described,
the NRCQM theory should be defined on a plane wave background. The properties
given by Eqs.(\ref{H6}) to (\ref{S6}) calculated in the gravity dual system are the predictions
for the corresponding thermodynamical quantities in the quantum mechanical theory in three spatial
dimensions. Its Hamiltonian should be the harmonic oscillator Hamiltonian described in the introduction
for $d=3$. We can also mention what happens if the other two rotation parameters in the general
metric of Gibbons et al \cite{Gibbons:2004ai} take non-vanishing finite values. In that case, they would be associated
with the angular momentum corresponding to rotations in two perpendicular planes of the three dimensional space.
In this case the Hamiltonian should describe the motion of a non-relativistic particle in the transverse
space in the presence of a magnetic field.

\subsection{DLCQ of Kerr-AdS$_7$ black hole}

For $d=4$, this black hole reduces to the seven dimensional case studied by MMT.
The eleven dimensional asymptotic metric is AdS$_7$$\times S^4$. For completeness we very briefly
discuss this case below.

The area of the event horizon is $A=\pi^3 \, r_+^3 \, (r_+^2+a^2)/\Xi$. The surface gravity
$\kappa$ is given by
\be
\kappa= r_+ (1+r_+^{2})\left(\frac{1}{(r_+^2+a^2)}+\frac{2}{r_+^2}\right)-\frac{1}{r_+} \, .
\ee
The Hawking temperature has been defined in the previous sections.
The angular velocity relative to a non-rotating frame at infinity is
\be
\Omega= \frac{a(1+r_+^2)}{r_+^2+a^2} \, .
\ee
The energy and angular momentum are \cite{Gibbons:2004ai}
\bea
E=\frac{\pi^2 m}{4(1-a^2)} \left(\frac{1}{(1-a^2)} +2 - \frac{1}{2} \right)\, , \,\,\,\,\,\,\,\,\,\,\,
J=\frac{\pi^2 m a}{4(1-a^2)^2} \, .
\eea
The first law of thermodynamics reads as $\delta E = T \, \delta S + \Omega \, \delta J$,
which by using the above equations
for $T$, $\Omega$ and $J$, is an exact differential provided that the entropy is
\be
S=\frac{A}{4} \, .
\ee
After we perform the DLCQ of the metric of Eq.(\ref{metricKerrAdSd}) we get
Eq.(\ref{metricKerrAdSdDLCQ}) for $d+3=7$
\bea
ds^2_{7-DLCQ} &=& \frac{(r^2+\sin^2\theta)}{(r^2+1)^2 - \frac{2 m}{r^2}} dr^2 - (1+r^2\sin^2\theta)(dx^+)^2
                  -\lambda (1+r^2) \cos^2\theta \,dx^+ \,dx^-  \nonumber \\
              & & + \frac{(r^2+\sin^2\theta)}{\cos^2\theta} d\theta^2
                  + \frac{m}{r^2} \frac{(-2 dx^+ + (dx^+-\lambda dx^-) \cos^2\theta)^2}{2(r^2+\sin^2\theta)}
                  \nonumber \\
              & &  + r^2 \sin^2\theta d\Omega^2_{3}  \, . \label{metricKerrAdS7DLCQ}
\eea
Now, we calculate the Hamiltonian $H$ and $N$ and obtain
\bea
&& H = -P_+ = \frac{r^-}{2 R^2} \, (E-J) = \frac{1}{8} \, \pi^2 \, r_+^2 \, (r_+^2+1)^2 \, r^- \, \lambda \, , \label{H7} \\
&& N = -P_- \, r^- = \frac{(r^-)^2}{(2 R^2)^2} \, J = \frac{1}{32} \, \pi^2 \, r_+^2 \, (r_+^2+1)^2 \, (r^-)^2 \, \lambda^2 \label{N7} \, .
\eea
These expressions have been obtained in the $a\rightarrow 1$ limit. $m$ has been set using the
condition of vanishing $\Delta_r(r_+)=0$ at the outer horizon leading to
$m=(r_+^2+a^2)(r_+^2+1) r_+^2/2$.
In this limit the temperature, the chemical potential and the entropy of the system are respectively
\bea
&& T = \frac{3 r_+^2 + 1}{2 \pi r_+} \, , \label{T7} \\
&& \mu = \frac{1}{r^- \lambda} \, \frac{(r_+^2-1)}{(r_+^2+1)} \, , \label{mu7} \\
&& S = \frac{1}{8} \, \pi^3 \, \lambda \, r_+^3 \, (1+r_+^2) \, r^- \, . \label{S7}
\eea
We have explicitly checked that the first law of thermodynamics
$\delta H=T \,\, \delta S-\mu \,\, \delta N$
is satisfied using the expressions (\ref{H7}-\ref{S7}). The radius of the circle along
$x^-$ at the horizon in units of the Planck length is
\be
R^-_{physical}=r^- \, \lambda \, \frac{(r_+^2+1) \, \cos^2\theta}{2 \sqrt{(r_+^2+\sin^2\theta})} \, R_{AdS_7} \, .
\ee
This radius shrinks at $\theta=\pi/2$ at any value of $r$.
However, the same discussion as in the previous cases applies, so that
there are no large contributions to the thermodynamic properties
from this region of the metric. Thus it is possible to assume
that this region of the metric can be ignored.
In this case the four dimensional NRCQM theory has been extensively discussed
in \cite{Maldacena:2008wh}.

\section{DLCQ of the BTZ black hole}

This case is interesting also because it corresponds to $d=0$.
We consider an alternative metric for the BTZ black hole given by Hawking et al \cite{Hawking:1998kw}
\be
ds^2_{BTZ} = -\frac{\Delta_r}{r^2} \left(dt - \frac{a}{\Xi} d\phi \right)^2 +
             \frac{r^2}{\Delta_r} dr^2 + \frac{1}{r^2}
             \left(a dt - \frac{r^2+a^2}{\Xi} d\phi \right)^2  \, , \label{metricalternativeBTZ}
\ee
where we have used the definition
\be
\Delta_r = (r^2+a^2)(1+r^2) - 2 m r^2\, .
\ee
Now, we carry out the DLCQ of the metric of Eq.(\ref{metricalternativeBTZ}).
First we use the scaling given by Eqs.(\ref{t}-\ref{a})
so that again we keep $\lambda$ as a parameter which takes into
account the amount of momentum injected in the compact direction.
We consider the limit for $R\rightarrow \infty$ that leads to $a\rightarrow 1$. We take
the limit on these expressions
\be
\lim_{R\rightarrow\infty}\Delta_r = (r^2+1)^2 - 2 m r^2 \, , \,\,\,\,\,\,\,\,\,\,\,\,\,\,
\lim_{R\rightarrow\infty}\Xi = \frac{1}{R^2 \lambda} \, . \nonumber \\
\ee
In this limit the metric Eq.(\ref{metricalternativeBTZ}) becomes
\be
ds^2_{BTZ-DLCQ} = \frac{r^2 dr^2}{(r^2+1)^2 - 2 m r^2} - (dx_+)^2
                  -\lambda (1+r^2) \,dx^+ \,dx^-
                  + \frac{m}{2} (dx^+ + \lambda dx^-)^2 \, . \label{metricBTZDLCQ}
\ee
The BTZ black hole can be embedded in string theory.
There are black string backgrounds characterized by D1-brane and D5-brane
charges $Q_1$ and $Q_5$, respectively, and by the momentum density, which are solutions of
type IIB string theory on $K3$. Also, one can consider the case $T^4$.
These objects have a dual description as a $c = 6 Q_1 Q_5$ conformal field theory whose target
space is the symmetric product of $Q_1 Q_5$ copies of $K3$ \cite{Strominger:1996sh}. The near-horizon geometry
of the spacetime solutions is AdS$_3 \times S^3 \times K3$ (or $T^4$). One can perform a
compactification of the black string on a circle. This yields a five dimensional black hole.
This corresponds to the compactification of AdS$_3$ on a circle $S^1$ \cite{Hyun:1997jv}.
In the very-near horizon limit the metric reduces to the one of AdS$_2$ \cite{Strominger:1998yg}.
Strominger has shown that the dual representation of the very-near horizon string theory is a
DLCQ conformal field theory, which suggests a connection with the matrix model \cite{Banks:1996vh,Susskind:1997cw}.
More recently Guica, Hartman, Song and Strominger conjectured that
extreme Kerr black holes are holographically dual to a chiral two-dimensional conformal
field theory and they obtained its central charge \cite{Guica:2008mu}.

\section{Discussion and conclusions}

We have considered the DLCQ of Kerr-AdS black holes of general dimension $d+3$, which have $d+2$
spatial dimensions and one time dimension.
All these backgrounds asymptotically lead to AdS$_{d+3}$ spaces
with a plane wave boundary. They correspond to
NRCQM theories in $d$ spatial dimensions
on plane wave backgrounds, at finite temperature and finite chemical potential.

When one starts from an AdS$_{d+3}$ spacetime the dual field theory is a $d+2$ relativistic field
theory which has $d+2$ dimensional Poincar\'e symmetry, i.e. $d+1$ spatial dimensions and one time dimension.
From this metric it is possible to obtain a plane wave structure at the boundary.
One can get it after its Penrose limit is taken.  So, the resulting dual field theory is defined on a
plane wave background. We can heat the theory up leading to a thermal
AdS, however this background does not admit a discrete light cone quantization because in that case
it would have a null direction. Thus, the point is that one has to solve the singularity in the
metric. In general there are a few ways to do it, for example sometimes there are certain fields to play with in the
gravity background so that by turning them on in a suitable configuration it is possible
to make the background regular, in order to be able to carry out the AdS/CFT correspondence analysis.
Another way to get a regular background is to consider a system with a certain amount of angular
momentum. This is the procedure carried out here along the lines proposed in \cite{Maldacena:2008wh}.
Thus, the choice of Kerr-AdS$_{d+3}$ black hole metrics
is very suitable for this purpose since it does both jobs simultaneously, i.e. includes a non-vanishing
angular momentum and sets the system at finite temperature. On this new metric the DLCQ programme
can be carried out, leading to a metric which also goes asymptotically to a plane wave at the boundary.
The initial Poincar\'e symmetry becomes a $d$ dimensional Galilean symmetry plus the
generators of the dilatations and the special conformal transformations, $D$ and $C$, respectively,
leading to the Schr\"odinger group. This tells us that the dual theory should
be a non-relativistic conformal quantum mechanical theory.

If these gravity solutions admitted an uplifting to string theory or M-theory, the DLCQ of the
Kerr-AdS$_{d+3}$ black hole metric would become embedded in string theory or M-theory.
The internal manifold used for the uplifting would have isometries which should be reflected as certain
symmetries of the NRCQM theory. In this respect we have discussed very briefly the already known
uplifting of AdS$_4$ spacetime to eleven dimensional supergravity configurations such as AdS$_4 \times S^7$,
AdS$_4 \times  \tilde{S}^7$,  AdS$_4 \times Q^{1,1,1}$ and AdS$_4 \times N^{0,1,0}$.
Also we have considered the uplifting of AdS$_6$ spacetime to massive type IIA supergravity leading to
AdS$_6 \otimes S^4$. They correspond to the large $N_c$ limit of
supersymmetric Yang Mills theories in $2+1$ and $4+1$ dimensions,
at zero temperature and zero chemical potential. This perhaps might give some clues to find out the
NRCQM theories after the DLCQ has been carried out, possibly along the lines of \cite{Motl:2003rw}
and \cite{Lowe:2003qy}.

Another interesting point is the study of possible phase transitions in these systems.
It is possible to compare the free energies of the DLCQ of the Kerr-AdS$_{d+3}$ black hole metrics
with the corresponding ones of the thermal AdS$_{d+3}$ spaces and see that a Hawking-Page type of
phase transition occurs for $r_+=1$ in all the cases above. This gives a transition
temperature $T_c$ which depends upon the dimensionality of the system under study. For temperatures lower than
$T_c$ the thermal AdS is the stable phase while for temperatures higher than $T_c$ the DLCQ of the
Kerr-AdS black hole dominates. The thermal AdS metric can only be trustable if $x^-$ is taken to be
non-compact. Thus, for the systems above a phase transition similar to the Hawking-Page one
should take place at $T_c=\frac{d}{2 \pi}$.

Since we have explicitly calculated thermodynamical
properties of the Kerr-AdS$_{d+3}$ black hole backgrounds, by the means of the gauge/gravity duality,
we conjecture that these are predictions for the corresponding thermodynamical properties of their
dual non-relativistic conformal quantum mechanical theories. We have not studied the specific form
of the NRCQM theories which are dual to the backgrounds described above. However, we think that
the examples studied here offer the interesting possibility of approaching the description of certain strongly
coupled systems in one and three spatial dimensions at finite temperature and finite chemical
potential which enjoy symmetries generated by the Schr\"odinger group.

~

~

\centerline{\large{\bf Acknowledgments}}

~

We are grateful to Nicol\'as Grandi for collaboration in an early stage of this project.
We thank Juan Maldacena for useful comments.
This work has been supported in part by CONICET, the National Scientific
Research Council of Argentina.

\newpage

\subsection*{Appendix: DLCQ of higher dimensional Kerr-AdS black holes}

In this appendix we introduce a number of equations for higher dimensional Kerr-AdS black holes.
We split it in even and odd dimensional cases. We have taken the expressions for $\kappa$,
the entropy, the energy and the angular momentum from reference \cite{Gibbons:2004ai}, and
then we have calculated the corresponding expressions for $H$, the number of particles and
the chemical potential.

~

{\it DLCQ of Kerr-AdS$_{d+3}$ black holes with $d+3$ odd}

~

Let us start with the case of dimension odd. The surface gravity is given by
\be
\kappa= r_+ (1+r_+^{2})\left(\frac{1}{(r_+^2+a^2)}+\frac{(\frac{d+2}{2}-1)}{r_+^2}\right)-\frac{1}{r_+} \, .
\ee
The area of the event horizon is
\be
A= {\cal {A}}_{d+1} \, \frac{(r_+^2+a^2)}{\Xi} r_+^{(d+2)-3}\, ,
\ee
where ${\cal {A}}_{d+1}$ is the volume of the $d+1$-sphere
\be
{\cal {A}}_{d+1} = \frac{2 \pi^{(d+2)/2}}{\Gamma[(d+2)/2]}  \, .
\ee
The Hawking temperature is
\be
T=\frac{1}{\beta}=\frac{\kappa}{2\pi} \, .
\ee
The angular velocity relative to a non-rotating frame at infinity is
\be
\Omega= \frac{a(1+r_+^2)}{r_+^2+a^2} \, .
\ee
The energy and angular momentum are
\bea
E=\frac{m {\cal {A}}_{d+1}}{4 \pi(1-a^2)} \left(\frac{1}{(1-a^2)} + \frac{(d+2)}{2} -1 - \frac{1}{2} \right)\, , \,\,\,\,\,\,\,\,\,\,\,
J=\frac{m a {\cal {A}}_{d+1}}{4 \pi(1-a^2)^2} \, ,
\eea
where $m=(1+r_+^2)(a^2+r_+^2) r_+^{d-2}/2$.
The first law of thermodynamics reads as $\delta E = T \, \delta S + \Omega \, \delta J$,
which by using the above equations
for $T$, $\Omega$ and $J$, is an exact differential provided that the entropy is
\be
S=\frac{A}{4} \, .
\ee
Now, we consider the DLCQ limit using the scaling described in above and
calculate the Hamiltonian $H$ and $N$
\bea
&& H = -P_+ = \frac{r^-}{2 R^2} \, (E-J) = \frac{d m \pi^{d/2} r^- \lambda}{8 \Gamma[(d+2)/2]} \ \, , \label{Hdodd} \\
&& N = -P_- \, r^- = \frac{(r^-)^2}{(2 R^2)^2} \, J = \frac{m \pi^{d/2} (r^-)^2 \lambda^2}{8 \Gamma[(d+2)/2]}  \label{Ndodd} \, .
\eea
These expressions have been obtained in the $a\rightarrow 1$ limit. $m$ is set using the
condition of vanishing $\Delta_r(r_+)=0$ at the outer black hole horizon leading to
$m=(1+r_+^2)(a^2+r_+^2) r_+^{d-2}/2$.
In this limit the temperature, the chemical potential and the entropy of the system are respectively
\bea
&& T = \frac{(d+2) r_+^2 + d - 2}{4 \pi r_+} \, , \label{Tdodd} \\
&& \mu = \frac{1}{r^- \lambda} \, \frac{(r_+^2-1)}{(r_+^2+1)} \, , \label{mudodd} \\
&& S = \frac{\pi^{(d+2)/2} r_+^{d-1} (1+r_+^2) r^- \lambda}{4 \Gamma[(d+2)/2]}  \, . \label{Sdodd}
\eea
We have explicitly checked that the first law of thermodynamics
$\delta H=T \,\, \delta S-\mu \,\, \delta N$
is satisfied using the expressions (\ref{Hdodd}-\ref{Sdodd}). The radius of the circle along
$x^-$ at the horizon in units of the Planck length
\be
R^-_{physical}=r^- \, \lambda \, \frac{(r_+^2+1) \, \cos^2\theta}{2 \sqrt{r_+^2+\sin^2\theta}} \,  R_{AdS_{d+3}} \, .
\ee
This radius shrinks at $\theta=\pi/2$ at any value of $r$.
However, the same discussion as in the previous cases applies.
Therefore, there are not large contributions to the above described thermodynamic properties
from this region of the metric, so that we assume that this region of the metric can be ignored.

~

{\it DLCQ of Kerr-AdS$_{d+3}$ black holes with $d+3$ even}

~

Now, let us consider the case of dimension even. The surface gravity is given by\footnote{Notice
that this expression differs by a sign with respect Eq.(4.7) of Gibbons et al. We have checked
that with this sign it reproduces the expressions for $d+3=4$ and also that in this form the first
law of thermodynamics is satisfied.}
\be
\kappa= r_+ (1+r_+^{2})\left(\frac{1}{(r_+^2+a^2)}+\frac{(\frac{(d+1)}{2}-1)}{r_+^2}\right)-\frac{1-r_+^2}{2 r_+} \, .
\ee
The area of the event horizon is
\be
A= {\cal {A}}_{d+1} \, \frac{(r_+^2+a^2)}{\Xi} r_+^{(d+1)-2}\, ,
\ee
where ${\cal {A}}_{d+1}$ is the volume of the $d+1$-sphere
\be
{\cal {A}}_{d+1} = \frac{2 \pi^{(d+2)/2}}{\Gamma[(d+2)/2]}  \, .
\ee
The Hawking temperature is
\be
T=\frac{1}{\beta}=\frac{\kappa}{2\pi} \, .
\ee
The angular velocity relative to a non-rotating frame at infinity is
\be
\Omega= \frac{a(1+r_+^2)}{r_+^2+a^2} \, .
\ee
The energy and angular momentum are
\bea
E=\frac{m {\cal {A}}_{d+1}}{4 \pi(1-a^2)} \left(\frac{1}{(1-a^2)} + \frac{(d+1)}{2} -1 \right)\, , \,\,\,\,\,\,\,\,\,\,\,
J=\frac{m a {\cal {A}}_{d+1}}{4 \pi(1-a^2)^2} \, ,
\eea
where $m=(1+r_+^2)(a^2+r_+^2) r_+^{d-2}/2$.
The first law of thermodynamics reads as $\delta E = T \,\, \delta S + \Omega \,\, \delta J$,
which by using the above equations
for $T$, $\Omega$ and $J$, is an exact differential provided that the entropy is
$S=\frac{A}{4}$.

Now, we consider the DLCQ limit using the scaling described in above and
calculate the Hamiltonian $H$ and $N$
\bea
&& H = -P_+ = \frac{r^-}{2 R^2} \, (E-J) = \frac{d m \pi^{d/2} r^- \lambda}{8 \Gamma[(d+2)/2]} \ \, , \label{Hdeven} \\
&& N = -P_- \, r^- = \frac{(r^-)^2}{(2 R^2)^2} \, J = \frac{m \pi^{d/2} (r^-)^2 \lambda^2}{8 \Gamma[(d+2)/2]}  \label{Ndeven} \, .
\eea
These expressions have been obtained in the $a\rightarrow 1$ limit. $m$ is set using the
condition of vanishing $\Delta_r(r_+)=0$ at the outer black hole horizon leading to
$m=(1+r_+^2)(a^2+r_+^2) r_+^{d-2}/2$.
In this limit the temperature, the chemical potential and the entropy of the system are respectively
\bea
&& T = \frac{(d+2) r_+^2 + d - 2}{4 \pi r_+} \, , \label{Tdeven} \\
&& \mu = \frac{1}{r^- \lambda} \, \frac{(r_+^2-1)}{(r_+^2+1)} \, , \label{mudeven} \\
&& S = \frac{\pi^{(d+2)/2} r_+^{d-1} (1+r_+^2) r^- \lambda}{4 \Gamma[(d+2)/2]}  \, . \label{Sdeven}
\eea
We have explicitly checked that the first law of thermodynamics
$\delta H=T \,\, \delta S-\mu \,\, \delta N$
is satisfied using the expressions (\ref{Hdeven}-\ref{Sdeven}). The radius of the circle along
$x^-$ at the horizon in units of the Planck length has the same expression as in the odd dimensional case.

~

{\it Asymptotic form of the DLCQ of Kerr-AdS black hole in arbitrary dimensions}

~

Applying the coordinate transformation (\ref{newcoord}) to the metric(\ref{metricKerrAdSdDLCQ}) it can be
recast as follows
\bea
d{\tilde s}^2_{d+3-DLCQ} &=& \left(1-\frac{2 \, m}{(1+\lambda \, \rho^2)^{(d+2)/2}} \,
                  \frac{\lambda^{(d+2)/2}}{y^{d+2}} \right)^{-1} \,
                  \frac{dy^2}{y^2} \nonumber \\
              & &   + y^2 (-dx^+ \, dx^- - \rho^2 \, (dx^+)^2 + d\rho^2 + \rho^2 \, d\Omega^2_{d-1}) \nonumber \\
              & & - (dx^+)^2 + \frac{m \, \lambda^{d/2}}{y^d}
              \frac{((1+2\lambda\rho^2)dx^++\lambda dx^-)^2}{2(1+\lambda \rho^2)^{(d+4)/2}} + \cdot \cdot \cdot \, ,
                  \label{metricKerrAdSdDLCQrhoy}
\eea
where $\cdot \cdot \cdot$ indicates subleading terms, including deformations of the plane wave
boundary metric proportional to $d\rho^2$, and other deformations
which come from the term of the metric (\ref{metricKerrAdSdDLCQ}) which is proportional to $m$.
The asymptotic behavior of this metric is similar to the metric (\ref{metric0}).

~

\end{document}